\documentclass[12pt,titlepage]{utarticle}

\usepackage{xcolor}
\usepackage{graphicx}
\usepackage{multirow}
\usepackage{amsfonts}
\usepackage{amssymb}
\usepackage{amsmath}
\usepackage{slashed}
\usepackage{CJK}
\numberwithin{equation}{section}

\newcommand{\be}{\begin{equation}}
\newcommand{\ee}{\end{equation}}

\def\a{\alpha}

\def\d{\delta}

\def\s{\sigma}

\def\L{{\cal L}}

\newcommand{\shgsTWICE}[1]{}
\def\frac#1#2{{{{#1}}\over{{#2}}}}
\def\hatco{\co}
\def\DeltaIZED{\Delta}

\usepackage{wasysym}

\newsavebox{\ns}
\newsavebox{\dbrane}
\newsavebox{\dbshort}

\usepackage{epsfig}

\renewcommand{\theequation}{\arabic{section}.\arabic{equation}}

\def\appendix{{\newpage\section*{Appendix}}\let\appendix\section%
        {\setcounter{section}{0}
        \gdef\thesection{\Alph{section}}}\section}
\def\theequation{\thesection.\arabic{equation}}

\newcommand\ba{\begin{eqnarray}}
\newcommand\ea{\end{eqnarray}}

\def\Dslash{\,\,{\raise.15ex\hbox{/}\mkern-12mu D}}
\def\Dbarslash{\,\,{\raise.15ex\hbox{/}\mkern-12mu {\bar D}}}
\def\delslash{\,\,{\raise.15ex\hbox{/}\mkern-9mu \partial}}
\def\delbarslash{\,\,{\raise.15ex\hbox{/}\mkern-9mu {\bar\partial}}}
\def\pslash{\,\,{\raise.15ex\hbox{/}\mkern-9mu p}}
\def\calDslash{\,\,{\raise.15ex\hbox{/}\mkern-12mu {\cal D}}}

\newcommand{\hh}{{1\over 2}}

\renewcommand{\ll}{_}
\newcommand{\uu}{^}
\def\ppp{\partial}
\renewcommand{\exp}[1]{{\rm exp}\{#1\}}

\newcommand{\m}{\mu}

\renewcommand{\dag}{{}^\dagger{}}

\renewcommand{\m}{\mu}
\newcommand{\n}{\nu}

\newcommand{\sqd}{^2}

\renewcommand{\hh}{{1\over 2}}

\newcommand{\eee}[1]{\ba{#1}\ea}
\renewcommand{\th}{\theta}

\newcommand{\llsk}{\hskip .5in}

\newcommand{\st}{{}^*}

\newcommand{\IZ}{\relax\ifmmode\mathchoice
{\hbox{\cmss Z\kern-.4em Z}}{\hbox{\cmss Z\kern-.4em Z}}
{\lower.9pt\hbox{\cmsss Z\kern-.4em Z}} {\lower1.2pt\hbox{\cmsss
Z\kern-.4em Z}}\else{\cmss Z\kern-.4em Z}\fi} \font\cmss=cmss10
\font\cmsss=cmss10 at 7pt
\newcommand{\inbar}{\,\vrule height1.5ex width.4pt depth0pt}
\newcommand{\IC}{{\relax\hbox{$\inbar\kern-.3em{\rm C}$}}}
\newcommand{\IQ}{{\relax\hbox{$\inbar\kern-.3em{\rm Q}$}}}
\newcommand{\IP}{\relax{\rm I\kern-.18em P}}

\renewcommand{\k}[1]{{k_{#1}}}

\newcommand{\ed}{\dot{e}}

\renewcommand{\k}{\kappa}

\newcommand{\cc}{{\cal C}}

\renewcommand{\cc}{{c_1}}

\renewcommand{\cc}{c}
\newcommand{\thb}{\bar{\theta}}

\newcommand{\cl}{{\cal L}}

\newcommand{\ald}{{\dot{\alpha}}}

\newcommand{\IR}{\relax{\rm I\kern-.18em R}}
\def\blfootnote{\xdef\@thefnmark{}\@footnotetext}

\newcommand{\bm}{\begin{matrix}}
\renewcommand{\em}{\end{matrix}}

\newcommand{\lba}{\left |}
\newcommand{\rba}{\right |}

\newcommand{\co}{{\cal O}}

\newcommand{\rr}[1]{(\ref{{#1}})}
\newcommand{\bbb}{\ba\begin{array}{c}}
\renewcommand{\eee}{\nonumber\end{array}\ea}
\newcommand{\een}[1]{\label{#1}\end{array}\ea}

\newcommand{\prpr}{^{\prime\prime}{}}

\def\hilo{{}_{{}_{{}_{{}_{{}_{}}}}} {}^{{}^{{}^{}}}}

\newcommand{\heading}[1]{\begin{center}\it {#1} \rm \end{center}}

\def\lrdd{\left ( ~}
\def\rrdd{\hilo \right )}
\def\lsqq{\left [ ~}
\def\rsqq{\hilo \right ]}

\def\bi{\begin{itemize}}
\def\ei{\end{itemize}}

\def\ed{\end{document}}

\def\cc{{\cal C}}

\renewcommand{\rr}[1]{(\ref{#1})}

\def\cc{\,}

\def\cc{\,}

\def\rmt{^{\rm T}}

\def\DDB{\bar{D}}

\def\dim{D}

\def\seff{S_{\rm eff}}
\def\leff{{\cal L}_{\rm eff}}

\def\ssqd{^{{\rm sqd}}}
\def\ssqd{^2}
\def\bbx{\square}
\def\pch{{{\bf P}_\chi}}
\def\pach{{{\bf P}_{\bar{\chi}}}}
\def\DDB{\bar{D}}
\def\DDb{\bar{D}}
\def\prs{{{\bf P}_s}}
\def\prt{{{\bf P}_{\perp}}}
\def\str{{\tt str}}
\def\istr{{\tt Istr}}
\def\k{\kappa}
\def\kb{\bar{\kappa}}
\def\istrz{{\tt Istr}_0}
\def\bigp{\hat{\IP}}

\begin{document}\begin{CJK*}[dnp]{JIS}{min}
\baselineskip=15.5pt
\preprint{IPMU-12-0090}
\title{The One-Loop Effective K\"ahler Potential.  \\[4mm] I~:~ Chiral Multiplets.
}
\author{Raphael Flauger
   \address{
      School of Natural Sciences,\\
      Institute for Advanced Study,\\
      Princeton, NJ 08540, USA\\
   }$^,$
   \address{
      Center for Cosmology and Particle Physics, \\
      Department of Physics,\\
      New York University\\
      New York, NY, 10003, USA\\
   }
     , Simeon Hellerman $^3$, \\Cornelius Schmidt-Colinet $^3$,
     and Matthew Sudano
   \address{
      Kavli IPMU, The University of Tokyo \\
      Kashiwa, Chiba  277-8583, Japan\\
   }
}
\def\kkka{ \hbox{ カ　}}
\def\ka{ \hbox{ カ　}}
\Abstract{We derive a universal formula for the one-loop renormalization of the
effective K\"ahler potential that applies to general supersymmetric
effective field
theories of chiral multiplets, with arbitrary 
interactions respecting ${\cal N} = 1$ supersymmetry in four dimensions.
The resulting expression depends only on the
tree-level mass spectrum and the form of the regulator.
This formula  
simplifies and generalizes 
existing results in the literature. We include two examples to illustrate its use. \vskip -5mm}
\maketitle

\tableofcontents
\newpage

\def\ka{カ}


\section{Introduction}

The last two decades have seen much progress in the understanding
of quantum corrections to supersymmetric field theories with four
supercharges. This is mostly attributable to the highly constrained 
behavior of BPS terms in the effective action -- in particular to the 
holomorphic superpotential, where certain non-renormalization properties 
have been demonstrated both in perturbative
\cite{Grisaru:1979wc} as well as in non-perturbative \cite{Seiberg:1993vc} 
calculations, and in various contexts in string theory and 
supergravity (see {\it e.g.} \cite{Dine:1986vd, Ellis:1987yx, seibergemdualitylectures}
and references therein).

On the other hand, the effective K\"ahler potential generally
receives quantum corrections. The perturbative one-loop contribution 
was the subject of many studies in a variety of cases,
as {\it e.g.} in the superspace calculation of \cite{Buchbinder:1994iw} for the 
Wess-Zumino model, and in \cite{Clark:1997ri} for nonlinear sigma 
models. A more systematic approach was carried out in \cite{Grisaru:1996ve}
for renormalizable theories in $D=4$, and generalized in
\cite{Brignole:2000kg} to the non-renormalizable case, albeit
still excluding supersymmetric higher-derivative terms.
The importance of the latter was emphasized {\it e.g.} in 
\cite{Antoniadis:2007xc}, both as generic features of the
(Wilsonian) low-energy effective action as well as in the context
of some supersymmetry breaking scenarios.

The purpose of the present work is to develop a theory 
of the renormalization of the effective K\"ahler potential at one loop
which is general enough to tackle generic effective field theories and 
still powerful enough to allow simple calculation even in complicated
situations.

We think that there are reasons for which such
a general approach is desirable. 
The K\"ahler potential can transmit CP- and flavor-breaking effects in realistic candidate models of supersymmetric particle physics that are tiny but nonetheless highly constrained experimentally.
Controlling the renormalization of the K\"ahler potential in effective theories is therefore
an essential but often tedious task, which our formula simplifies.

\def\phenoCUTOUT{Many of the sharpest constraints on
supersymmetric model building have to do with the preservation of
approximate global symmetries, such as continuous approximate
flavor symmetries, baryon number, lepton number,
or else discrete symmetries such as CP or R-parity which is closely related
to baryon number conservation.  Often, these symmetries are known
to be approximately or exactly conserved by the superpotential terms,
as these terms are directly related to observable Yukawa couplings in
the Standard Model after SUSY breaking. 

These symmetries may become approximate through the
details of the theory at high energies. There they may actually be broken
by effects of a UV complete theory of gravity, which are realized
in lower energies as explicit non-renormalizable terms in the
low-energy Lagrangian (this could for example occur in gravity 
mediation models). If one has to decide which UV theory
is a viable candidate by that it preserves the approximate
global symmetries at low energies to an experimantally tolerable level,
one might be tempted to resort to perturbation theory;
the large hierarchy between the Planck scale and the SUSY
breaking scale reduces the complexity of quantum effects due to
non-renormalizable couplings, while weak gauge couplings near the
fundamental scale suppress the coefficients of higher loops due to 
renormalizable interactions (of course, dimensional suppression of 
nonrenormalizable terms is not quite enough: generic flavor- and 
CP-breaking terms in the K\"ahler potential will give rise to FCNC 
effects and observable neutron EDM, for instance, that significantly 
violate experimental bounds).
However, even if we stick to the perturbative treatment,
the problem of tracking symmetry-breaking
nonrenormalizable terms in realistic models is generically involved 
even at tree level, with the complexity of the problem growing rapidly 
if quantum corrections are taken into account. }

In the hope that ideas along these lines may help
winning the reader's approval for a general treatment
of the quantum correction $\Delta K$ of the K\"ahler potential, 
let us proceed by stating our result. We shall show 
that, for any given regulator, there is a simple 
formula for $\Delta K$ at one loop that reduces 
to a sum over on-shell degrees of
freedom running in the loop, with each massive
degree of freedom contributing an amount depending only on its
mass-squared.  We refer to this property by saying that the
formula for the one-loop renormalization of $K$ by virtual degrees of
freedom is purely \it mass-dependent. \rm  By this we mean to emphasize
that none of the three- or four-point couplings affects the one-loop value of
$\Delta K$, except through their influence on the physical masses of the
linearized modes.   

The unfamiliar
aspect of the mass-dependent formula is that unphysical degrees of freedom, both
positive and negative norm poles of the propagators that lie above the
cutoff of the theory, contribute
democratically with the physical degrees of freedom whose
masses lie below the cutoff.  We will show that there is no contradiction
here, so long as we are working in a self-consistent
regime of the application of
the effective field theory. 

 In the present paper, we restrict ourselves to
the case where the degrees of freedom circulating in the loop consist
of chiral multiplets only.  When massive
vector multiplets are present, there is a
similar mass-dependent formula.  Due to gauge invariance, there are
subtleties involved in its derivation and interpretation, so for clarity of presentation we will defer the discussion of vector multiplets to a subsequent paper.
For both chiral and vector multiplets, the contribution of
massless multiplets is \it not \rm purely mass-dependent, but in both
cases can be derived by taking an $m\to 0$ limit of the
corresponding mass-dependent
formul\ae~for massive degrees of freedom.

The organization of this paper is as follows.  After fixing some notation in section
\ref{sec:supertrace}, we
compute in section \ref{secthree} the one-loop effective K\"ahler potential for
a general effective theory of a single chiral superfield in four dimensions with
${\cal N} = 1$ supersymmetry,
defined at the quantum level with
a general supersymmetric regularization procedure.  We then compare the result
to the spectrum of linearized solutions to the equations
of motion and find that the one-loop correction to the K\"ahler potential is always of the form
\begin{equation}\label{lastformula}
\boxed{\begin{array}{ccc}&&\\[-15pt]
&
\displaystyle{\Delta K = 
\frac{\hbar}{{2 \cc (2\pi)\uu D}}\cc 
\int  \cc 
\frac{d\uu D P}{P\sqd} \sum\ll{j = 1}\uu J
\lsqq \cc {\rm ln} (P\sqd +  \m\ll j \sqd) - {\rm ln}(\m\ll j\sqd) \cc \rsqq}\,.
&\\[-15pt]
&&
\end{array}
}
\end{equation}
The sum runs over all supermultiplets of on-shell excitations with masses $\mu_j$, including unphysical super-cutoff solutions to the equations of motion
with masses lying far above the cutoff. However, in minimal subtraction schemes such as $\overline{\text{DR}}$ the contributions from the unphysical super-cutoff solutions vanish if equation~\eqref{lastformula} is properly interpreted, and the sum can be restricted to the light modes. 
In section \ref{secfour} we show that the validity of equation~\eqref{lastformula} extends to the case of an arbitrary number of chiral superfields.
In section \ref{secfive} we consider examples in four and two dimensions to shed light on the regularization issue. Appendix~\ref{AppendixWBconventions} summarizes our conventions. Finally, Appendix~\ref{AppendixOproperties} is devoted to several useful identities that were used in the derivation of~\eqref{lastformula}.

\section{Notation for supertraces and superdeterminants}\label{sec:supertrace}%
\setcounter{equation}{0}

For operators  ${\cal O}$ acting on the ring of functions on superspace,
the supertrace is defined as
\be\label{str}
{\rm str}({\cal O}) \equiv \cc {\rm tr}
\lrdd (-1)\uu F\cc {\cal O}\cc \rrdd
= {\rm tr}\lrdd {\cal O}\cc (-1)\uu F \cc \rrdd\ ,
\ee
where $F$ is the fermion number operator.
With the supertrace we define the superdeterminant (Berezinian)
in the usual way as
\be\label{sdet}
{\tt sdet}({\cal O}) = \exp{{\rm str}\big(
{\rm ln}({\cal O}) \big)}
\ .
\ee
We shall work with various partial traces whose notation we 
will now introduce.
Our focus is on theories with four real supercharges in $D\leq4$ so that locally smooth functions on superspace form a space of
the structure $C^{\infty}(\mathbb{R}^D)\otimes{\cal V}$,
where ${\cal V}$ is a 16-dimensional graded vector space 
(built from all combinations of $\theta$s).
Restrictions of traces or determinants over only ${\cal V}$ will be denoted  by
a subscript (${\rm tr}_{\cal V}$, \it etc.\rm). In particular, the restriction of the
supertrace can be written in terms of the integral expression
\be\label{strintegral}
\str_{\cal V}(\co) = 16 \cc \displaystyle{\int} \cc d\ssqd\th\cc d\ssqd\thb\cc
d\ssqd \k \cc d\ssqd\kb\cc \exp{- \k\th - \kb\thb} \,
\co \,
\big[
\exp{+ \k\th + \kb\thb} \big]\ ,
\ee
where the $\k$ and $\kb$ are Grassmann parameters
that play the roles of possible ``eigenvalues'' of the
derivatives $D_{\alpha}$ and $\bar{D}_{\dot{\alpha}}$. The square brackets
mean that $\co$ is evaluated on the exponential contained in
the brackets. The demonstration
that this formula holds true can be done straightforwardly by choosing a basis
for the operators on the 16-dimensional graded vector space.

We will also need to take the supertrace 
over chiral and antichiral superfields alone\footnote{This is the primary difference between our method and that of \cite{Grisaru:1996ve}, where instead of performing this sort of chirally projected path integral, chiral fields were replaced with unconstrained superfields $\Phi=\bar D^2\Psi$ (and ghosts that don't contribute at leading order).}. For this
we define the projection operator 
\be\label{chiralprojectors}
\bigp= \lrdd\bm \pch & 0 \cr 0 & \pach \em \rrdd\quad\,,
\ee
where $\pch$ and $\pach$ are the projectors
onto chiral and antichiral superfields 
 $\rr{pchdefmemo} , \rr{pachdefmemo}$, respectively.
The ``constrained supertrace'' over chiral and antichiral
superfields shall then be
\be\label{cstr}
{\tt Cstr}(\hatco) = {\tt str}(\bigp\cc \hatco)
= {\tt tr}((-1)\uu F\cc \bigp\cc \hatco)\ ,
\ee
with the constrained superdeterminant
\be\label{csdetdefmemo}
{\tt Csdet}(\hatco) = \exp{{\tt str}\big(
\bigp \cc {\rm ln}(\hatco) \big)
}
= \exp{{\tt tr}\big(
(-1)\uu F \cc \bigp \cc {\rm ln}(\hatco) \big)
}\ .
\ee
In an analogous way we define the constrained {\it trace}
of the operator $\mathcal{O}$ in the ordinary case as 
\be\label{ctr}
 {\tt Ctr}(\hatco) \equiv {\tt tr}(\bigp\cc \hatco)\ ,
\ee
and its constrained determinant as
\be\label{cdet}
{\tt Cdet}(\hatco) = \exp{{\tt ctr}\big(
{\rm ln}(\hatco) \big)
} = \exp{{\tt tr} \big(
\bigp \cc {\rm ln}(\hatco) \big)
}\ .
\ee

The advantage of writing the supertrace as in \rr{strintegral} is that 
it gives us a quantity which yields the supertrace upon integration 
over the Grassmann parameters $\th$ and $\thb$. We define
\be\label{Istr}
\istr_{\cal V}(\co) \equiv 16 \cc \displaystyle{\int} \cc d\ssqd \k\cc  d\ssqd\kb \cc \exp{- \k\th - \kb\thb} \,
 \co \,\big[
\exp{+ \k\th + \kb\thb} \big]
\, .
\ee
In the next section we will consider the particular case of an operator $\mathcal{O}$
which is 
a superfield, and hence completely defined if we know its value 
at $\theta=\bar{\theta}=0$. A straightforward calculation shows
that the $\theta$-integrand of the supertrace  at $\theta=\bar{\theta}=0$
is then given by
\be\label{istrzformulamemo}
\istr_{{\cal V}|0}(\co) = \co\,\big[ 
\th\ssqd \thb\ssqd \big]
\ll{\th = \thb = 0}\, .
\ee
For a translationally invariant operator $\mathcal{O}$ we can split 
the trace into traces over spaces $E_P$ of fixed momentum $P$
and write
\begin{align}\label{strfixedP}
\str(\co)&={{V\ll \dim}\over{(2\pi)\uu\dim}} \cc \displaystyle{\int} dE\cc d\uu {\dim - 1} P\cc
\str\ll{E_P\otimes{\cal V}} (\co)
\\\nonumber
&=16 \cc 
{{V\ll \dim}\over{(2\pi)\uu\dim}} \cc\!\!\! \int\! dE\cc d\uu {\dim - 1} P\cc \cc d\ssqd\th\cc d\ssqd\thb\cc
d\ssqd \k \cc d\ssqd\kb\cc \exp{- \k\th - \kb\thb} \,
\co(P) \,\big[
\exp{+ \k\th + \kb\thb} \big]
\end{align}
and so on, where ${\cal O}(P)$ stands for the operator restricted to the eigenspace
of eigenvalue $P$ and $V_D$ is the space-time volume. 
Throughout the calculation we will assume that the volume and the 
integral are regularized in some way. In fact, different regularization prescriptions
will have different effects eventually. For the sake of clarity we will,
however, in the following two sections ignore the ambiguities of
regularization. We shall come back to this in section~\ref{secfive}.

\section{Loop correction from a single chiral superfield}\label{secthree}
In this section we derive the one-loop effective K\"ahler potential in the form~\eqref{lastformula} for a general $\mathcal{N}=1$ supersymmetric effective field theory of a single chiral superfield. We begin by presenting a convenient formula for the one-loop effective K\"ahler potential in terms of the integrand of a supertrace in subsection~\ref{gen}. By relating the result of an explicit evaluation of this formula to the spectrum of tree-level masses, we then derive equation~\eqref{lastformula} in subsection~\ref{sec:ChirSFCalc}.

\subsection{General remarks}\label{gen}

For a given background  $\Phi ,\, \Phi\dag$ 
let ${\cal O}$ be the operator defining the quadratic action for fluctuations
$\chi \equiv \d \Phi,\, \chi\dag \equiv \d \Phi\dag$,
\be\label{quadraticaction}
S\ll{\rm quad} = \hh\cc
\displaystyle{\int} d\uu {\dim+4} z \cc 
\left(\,\chi\dag,\chi\,\right)\,
{\cal O} \,
\dbinom{\chi}{\chi\dag}
\, .
\ee
In this formula, $d^{D+4}z$ is the measure on superspace, 
and we have used the fact that for chiral superfields an integral 
over half of the fermionic coordinates can be written in terms of a 
full superspace integration after including appropriate projectors.
Notice that the operator 
${\cal O}$ as defined is Hermitian if the action is real.
Moreover, since we assume that the underlying theory is supersymmetric, it
is constructed only out of the operators $\hat{P}, D, \DDB,$ and the background
superfields $\Phi$. We will come back to the form of ${\cal O}$ more 
explicitly in section \ref{sec:ChirSFCalc}.

The gaussian path integral over fluctuations $\chi,\chi\dag$ then yields
\be
Z\ll{\rm gaussian} = \lsqq\cc {\tt Csdet }({\cal O}) \rsqq\uu{-\hh}\,.
\ee
For a partition function $Z$, the effective action is given by
\be
\seff = - i \hbar\cc {\rm ln}(Z)\,,
\ee
and extracting the purely one-loop contribution $\Delta S$ to the effective action in the fixed background $\Phi,\Phi\dag$ amounts to including only the determinant
over Gaussian fluctuations, such that
\be
\DeltaIZED S  = -i \hbar \cc {\rm ln}(Z\ll{\rm gaussian})
=  {{i\hbar}\over 2}\cc {\tt Cstr}\big(
{\rm ln}({\cal O}) \big)
= {{i\hbar}\over 2}\cc  {\tt str} \big(
\bigp \cc {\rm ln}({\cal O}) \big)
\ .
\ee

In order to evaluate the 1-loop corrections to the K\"ahler potential it is sufficient to consider supersymmetric (and translationally-invariant) background field configurations
\be\label{simpleansatzmemo}
D\Phi = \DDB\Phi\dag =
0\ ,
\ee
such that
$\Phi = \phi = {\rm const.}$ and $\Phi\dag = \phi\st = {\rm const}$.  
In this case, the value of the effective Lagrangian $\leff$
is independent of position, and we have
\be
S_{\rm eff} = V\ll\dim\cdot {\cal L}_{\rm eff}\,.
\ee
In particular, for translationally invariant backgrounds the operator ${\cal O}$ on Gaussian
fluctuations $\chi$ and $\chi\dag$ is also translationally invariant, and thus
the corresponding one-loop contribution to the effective Lagrangian is
\be
\DeltaIZED \cl =  {{i\cc \hbar}\over {2 V\ll\dim}}
\cc  {\tt str} \big(
\bigp \cc {\rm ln}({\cal O}) \big)
\,.
\ee
As ${\cal O}$ is translationally invariant, we decompose the trace into a basis of
eigenfunctions of $\hat{P}\ll\m$ with eigenvalues $P\ll\m$,
and write the formula for the one-loop effective Lagrangian as
\be
\DeltaIZED \cl  = {{ i\cc \hbar}\over {2 \cc (2\pi)\uu\dim }}
\cc \displaystyle{\int} d E \cc  d\uu {\dim - 1} P \cc
 {\tt str}\ll{E_P\otimes\cal V}\cc  \big(
 \bigp \cc {\rm ln}({\cal O}) \big)
 \,.
\ee
The one-loop contribution $\Delta{\cal L}$ to the Lagrangian is given by 
the contribution $\Delta K$ to the K\"ahler potential\,,
\be\label{knormalization}
\Delta {\cal L}= \displaystyle{\int} d\ssqd \th \cc d\ssqd \thb \cc \Delta K ,
\ee
which we can now express by means of the integrand 
of the supertrace as defined in section \ref{sec:supertrace}
as
\be\label{DeltaKExpressionEuc}
\DeltaIZED K (\phi, \phi\st) =  {{i\cc \hbar}\over {2 \cc (2\pi)\uu\dim }}
\cc \displaystyle{\int} dE\cc d\uu {\dim - 1} P\cc \istr\ll{E_P\otimes{\cal V}}\cc \big(
\bigp \cc {\rm ln}({\cal O}) \big)
\ .
\ee
In the next section we will parametrize the general operator ${\cal O}$ and
compute the value of $\istr(\bigp\,{\rm ln}{\cal O})$
in terms of that parametrization.

\subsection{Parametrization and calculation of the 1-loop K\"ahler potential}\label{sec:ChirSFCalc}

We divide the action for 
fluctuations $\chi = \d \Phi, \chi\dag = \d \Phi\dag$
into a half-superspace term, its complex conjugate, and the full-superspace term:
\be\label{Ldivided}
{\cal L} = \hh \int\! \cc d\ssqd \th \cc \chi \cc {\cal O}\ll{\rm half} \cc\chi
+ \hh \int\! \cc d\ssqd \thb \cc \chi\dag \cc {\cal O}\ll{\rm \overline{half}} \cc\chi\dag
+ \hh \int\! \cc d\ssqd \th \cc d\ssqd \thb \cc \big[
\chi {\cal O}\ll{\rm full} \chi\dag + \chi\dag {\cal O}\ll{\rm full}\rmt \chi \big]
\ .
\ee
We are working in a background that satisfies the ansatz \rr{simpleansatzmemo}, 
so in particular the background obeys Lorentz invariance.
It is important to notice that as a consequence of
Lorentz invariance, hermiticity of ${\cal O}$, and the reality of the action,
we lose no generality by taking
\begin{align}\label{sudanonormalformmemo}
{\cal O}\ll{\rm half}& =
A\ll 1(\bbx)\,,\qquad\quad
{\cal O}\ll{\rm \overline{half}} =
 A\ll 1\st(\bbx)\,,
\nonumber\\
{\cal O}\ll{\rm full} &= {\cal O}\ll{\rm full} \rmt =
 {\cal O}\st\ll{\rm full} = {\cal O}\ll{\rm full} \dag = A\ll 2(\bbx)\ .
\end{align}
Since the half-superspace integral 
\be
I \equiv \displaystyle{\int} d\ssqd \th \cc \chi {\cal O}_{\rm half} \chi
\ee
can always be written as a full superspace integral
\be
I =  \displaystyle{\int} d\ssqd \th \cc d\ssqd\thb\cc \chi \left(- {{D\ssqd}\over{4\bbx}}\cc {\cal O}_{\rm half}\right) \cc \chi\ ,
\ee
and likewise for antichiral superspace integrals
\be
\bar{I} \equiv  \displaystyle{\int} d\ssqd \thb \cc \chi\dag {\cal O}_{\rm \overline{half}} \chi\dag
 =   \displaystyle{\int} d\ssqd \th \cc d\ssqd\thb\cc \chi\dag \left(- {{\DDb\ssqd}\over{4\bbx}}\cc {\cal O}_{\rm \overline{half}}\right) \cc \chi\dag\ ,
\ee
the quadratic Lagrangian for fluctuations around the given background is in fact
\be\label{Lagdensity}
\cl = \hh \displaystyle{\int} d\ssqd \th \cc d\ssqd \thb \cc (\chi\dag,\chi) \cc
{\cal O} \cc \dbinom{\chi}{\chi\dag} 
\ee
with
\be\label{a12def}
{\cal O} = \left ( \bm  A\ll 2 (\bbx) &  ( - {{\DDb\ssqd}\over{4 \bbx}} ) \cc A\ll 1 \st (\bbx) 
\cr  (- {{D\ssqd}\over{4 \bbx}} ) \cc A\ll 1 (\bbx)  
& A\ll 2(\bbx) \em \right )\ .
\ee
From here on out, we will take the action to be ``local" -- that is, the operators
$A\ll 1$ and $A\ll 2$ are polynomials or at least approximable 
in all respects by polynomials in $\bbx = -P\sqd$. 
Let us collect a few observations about the operators $A\ll 1$ and $A_2$:
\bi
\item{$A\ll 2$ is real.}
\item{$A\ll 1$ need not
be real.}
\item{$A\ll 2$ can have an arbitrary dependence on the 
constant background fields $\phi,\phi^*$.
\item{The non-constant terms in $A\ll 1$
can have arbitrary dependence on background fields
$\phi, \phi^*$, while the constant
term in $A\ll 1$ must depend holomorphically on $\phi$. The constant
term in $A\ll 1(\bbx)$ is identified as $W\prpr(\phi)$, where $W$ is
the superpotential.}}
\ei
The operator ${\cal O}$ depends only on $P\sqd$ rather than on $P$ in
general, so the Wick rotation is particularly unproblematic; we will
simply replace the Lorentzian invariant $- E\sqd + P\ll{\rm spatial}\sqd$
with the Euclidean invariant $P\sqd$.

Starting from \rr{DeltaKExpressionEuc}, we defer
the calculation to the Appendix, and take the results 
\rr{bigplogcaloval} for the logarithm of ${\cal O}$ and 
formula \rr{UsefulFormulaA} for $\str\ll {\cal V}$ of $\bigp$ in
order to write the one-loop correction as
\be\label{FinalFormula}
\Delta K(\phi,\phi^*) =  - {{\hbar}\over{2 \cc (2\pi)\uu D}}\cc 
\displaystyle{\int} \cc d\uu D P  \cc {\rm tr}_{E_P}\left(\frac1\bbx
\,{{\rm ln}\lrdd    A\ll 2\sqd - {{|A\ll 1|\sqd}\over\bbx } \rrdd }\right)\ .
\ee
As mentioned above, we will characterize $\Delta K$ in terms of a polynomial function of $P\sqd$  which we shall call the spectral polynomial $\s(P\sqd)$. It will have
the property that its roots lie at $-\m\sqd$, where $\m$ is a mass of the tree-level spectrum.
Our aim is to prove formula \eqref{lastformula} by relating $\s(P\sqd)$ to 
the polynomial appearing inside the logarithm of the momentum 
integrand in \rr{FinalFormula}.

So how do we define, and calculate, the spectral polynomial?  We would
like the roots of $\s(P\sqd)$ to encode the linearized
solutions to the equations of motion of the effective field theory,
with the correct multiplicities.
Written in superspace, these equations are 
\be
{\cal O} \,
\dbinom{\chi}{\chi\dag} = 0\ ,
\ee
and we can solve them on each eigenspace of $P$ separately. 
The tree-level excitations therefore lie exactly at the values of $P^2$ such that
${\cal O}$ has a vanishing eigenvalue at these specific momenta -- that is, when
its constrained determinant vanishes on the eigenspace $E_P$.
The constrained determinant
${\tt Cdet}$ rather than the ordinary determinant
is the
relevant object, because propagating modes of $\chi$ ($\chi\dag$) must
satisfy both the equations of motion and the chiral (antichiral) constraint.
Thus our definition for the spectral
polynomial is
\be\label{SpectralPolynomial}
\s(P\sqd):={\tt Cdet}_{E_P} (\sqrt{\bbx}\,{\cal O})\,.
\ee
The square root of $\bbx$ has been included
 in order to remove possible divergences if $P$ goes to 0.

Given our parametrization \eqref{a12def}, let us write 
the operator $\s(\bbx)$ 
in terms of the functions $A\ll 1(\bbx)$
and $A\ll 2(\bbx)$. 
In a derivation which we defer to Appendix \ref{AppendixOproperties},
we obtain in  \rr{logsigmaformula}
the expression
\be
\s(\bbx) = (\bbx)^4 \cc \lrdd A\ll 2(\bbx)\sqd - {{|A\ll 1|\sqd}\over\bbx} \rrdd^4\,.
\ee
Hence ${1\over 4} {\rm ln}(\s(\bbx))$
is the same, up to constant
terms, 
as  the operator in the trace which appears in the integrand in \rr{FinalFormula} :
\be\label{sigmaina12}
{\rm ln} \lrdd A\ll 2 \sqd - {{|A\ll 1|\sqd}\over\bbx } \rrdd
=
 - {\rm ln}(\bbx)  + {1\over 4} {\rm ln}\left( \s(\bbx)  \right) \ .
\ee
Let us consider how this expression is related to the spectrum of linearized modes.
The function $\s$ 
is a polynomial in $P\sqd$ of some finite order $4J$,
so we can write it in a factorized form as
\be\label{sigmadef}
\s(P\sqd) = N\cc\prod_{j=1}^{J}(P^2+\mu_j^2)^4\,.
\ee
Here the $\mu_j$ are mass parameters
of the linearized modes, and $N$ is a normalization constant. 
The parameter $j$ counts the distinct multiplets, which 
in an ${\cal N}=1$ theory form degenerate sets containing four 
states each. 
Both the $\mu_j$ and $N$ depend on the background fields $\phi$ and $\phi^*$.
Let us point out that $j$ in general runs over both
``physical states", {\it i.e.}  linearized states within the validity regime
of the original Lagrangian below some cut-off scale, as well as
``super-cut-off states" which lie beyond that scale. 
Evaluating \eqref{sigmadef} at $P = 0$ leads to the identity
\be\label{importantfactorizationeqmemo}
N\uu{1\over 4}  \cdot \prod\ll{j = 1}\uu J \m\ll j \sqd = \s(0)\uu{1\over 4}
 = \lba A\ll 1\rba\sqd\bigg|_{P\sqd = 0}\,.
\ee
Now, as we have observed, $A\ll 1|_{P\sqd = 0}$ is the second derivative of the superpotential. Therefore
\eqref{importantfactorizationeqmemo} is the norm square of a function which is holomorphic 
in the background field $\phi$.
If we use \eqref{importantfactorizationeqmemo} 
in \eqref{sigmadef}, we therefore find for the logarithm of $\s(P^2)$
\begin{align}\label{LogSigmaFormula}
{1\over 4} {\rm ln}(\s(P\sqd)) &= {1\over 4} \cc {\rm ln}(N) + \sum\ll{j = 1}\uu J
{\rm ln}
(P\sqd +  \m\ll j \sqd)
\nonumber\\
&= \displaystyle{\sum\ll{j = 1}\uu J}
\lsqq \cc {\rm ln} (P\sqd +  \m\ll j \sqd) - {\rm ln}(\m\ll j\sqd) \cc \rsqq\nonumber\\
&+ ({\rm HOLOMORPHIC}) + ({\rm ANTIHOLOMORPHIC})\,,
\end{align}
and \eqref{sigmaina12} hence yields the expression
\begin{align}\label{massdependentequation}  
{\rm ln}\lrdd    A\ll 2\sqd - {{|A\ll 1|\sqd}\over\bbx } \rrdd  \bigg|_{\bbx=-P^2}&= 
- {\rm ln}(-P^2) +  \sum\ll{j= 1}\uu J\lsqq \cc {\rm ln} (P\sqd +  \m\ll j \sqd) - {\rm ln}(\m\ll j\sqd) \cc \rsqq\nonumber\\
&+ ({\rm HOLOMORPHIC}) + ({\rm ANTIHOLOMORPHIC})
\end{align}
for the integrand in \eqref{FinalFormula}.
Since holomorphic and constant terms in the K\"ahler potential do
not contribute to the physical action and can be ignored, 
the final formula for $\Delta K$ is
\begin{equation}
\boxed{\begin{array}{ccc}&&\\[-15pt]
&
\displaystyle{\Delta K = 
 {{\hbar}\over{2 \cc (2\pi)\uu D}}\cc 
\int  \cc 
{{d\uu D P} \over{P\sqd}} \sum\ll{j = 1}\uu J
\lsqq \cc {\rm ln} (P\sqd +  \m\ll j \sqd) - {\rm ln}(\m\ll j\sqd) \cc \rsqq}\,.
&\\[-15pt]
&&
\end{array}
}
\end{equation}
We point out again that while the integral in
this formula is regularized, the sum over $j$
includes both physical and super-cut-off modes at the level of the integrand.
Different regularization procedures interact with this
feature in different ways. We will come back to this issue by 
considering concrete examples in section \ref{secfive}.

\section{Loop correction from multiple chiral superfields}\label{sec:multifield}\label{secfour}
\setcounter{equation}{0} \def\ka{ \hbox{カ}}

We now work with $n$ chiral superfields $\Phi\uu a$, $a=1,\ldots,n$.  Expanding
again around a given
vacuum, we collect the fluctuation fields $\chi\uu a$ into a vector $\chi$.  
Then the matrix defining the quadratic action 
\be
\cl = \hh \displaystyle{\int} d\ssqd \th \cc d\ssqd \thb \cc (\chi\dag,\chi) \cc
{\cal O} \cc \dbinom{\chi}{\chi\dag}
\ee
is
\be\label{coparam} 
{\cal O} = \left ( \bm {\cal O}_{\overline{\rm full}} &  ( - {{\DDb\ssqd}\over{ 4 \cc \bbx} } ) \cc  {\cal O}_{\overline{\rm half}}
\cr 
 (- {{D\ssqd}\over {4 \bbx} } ) \cc  {\cal O}_{\rm half}
 & {\cal O}_{\rm full}  \em \right )\,,
\ee
where now ${\cal O}_{\rm full},\, {\cal O}_{\overline{\rm full}},\, {\cal O}_{\rm half}$ 
and ${\cal O}_{\overline{\rm half}}$ are $n\times n$ matrices.
The reality of the action again imposes a hermiticity condition 
on the matrices,
\be
{\cal O}_{\overline{\rm full}}\dag = {\cal O}_{\overline{\rm full}} \ , 
\llsk {\cal O}_{\rm full}\dag = {\cal O}_{\rm full} \ , 
\llsk {\cal O}_{\rm half}\dag = {\cal O}_{\overline{\rm half}}\ ,
\ee
and the Bose statistics of the chiral superfields allow us to impose
\be
{\cal O}_{\overline{\rm full}}\rmt = {\cal O}_{\rm full} \ , \llsk 
{\cal O}_{\overline{\rm half}}\rmt = {\cal O}_{\overline{\rm half}} \ , \llsk 
{\cal O}_{\rm half}\rmt = {\cal O}_{\rm half}\ .
\ee
Taking also Lorentz invariance into account,  we can parametrize our matrix ${\cal O}$ as
\bbb
{\cal O} \equiv \left ( \bm A_2 (\bbx) &   - {{\DDb\ssqd}\over{ 4 \cc \bbx} }  \cc A_1\st (\bbx)
\cr 
 - {{D\ssqd}\over {4 \bbx} }  \cc  A_1 (\bbx)   
 & A_2\st (\bbx)  \em \right )\ ,
\een{Omultifield}
where $A_1$ is symmetric ($A_1\rmt = A_1$), and $A_2$ is hermitian.
With the ansatz \rr{simpleansatzmemo}, we have
\be\label{DeltaKExpressionLorNByN}
\DeltaIZED K (\phi, \phi\st)  = - {{ \hbar}\over {2 \cc (2\pi)\uu\dim }}
\cc \displaystyle{\int} d\uu \dim P\cc \istr\ll{E_P\otimes{\cal V}}\cc \big(
\bigp \cc {\rm ln}({\cal O}) 
\big)
\ee
in Euclidean signature, where now the supertrace includes a trace over the
$n\times n$ matrix structure acting on the $n$ chiral superfields and their conjugates.
Since it is constructed from superfields, the integrand is again a superfield as well,
and hence determined by its $\theta=\bar{\theta}=0$ component.
We will now proceed in a slightly different way than in section~\ref{sec:ChirSFCalc} and
represent ${\tt Istr}$, evaluated at $\th = \thb = 0$, as a trace, 
by using an additional operator insertion.
Indeed, for all operators on superspace of the form \eqref{Omultifield}, we can write
\begin{align}
&{\tt Istr}_{E_P\otimes{\cal V}|0}(\bigp \cc {\rm ln}\,{\cal O}) =
 {\tt tr}_{E_P\otimes{\cal V}|0}( \ka  \cdot \bigp \cc {\rm ln}\,{\cal O})\ ,\\
&{\rm where}\quad\hbox{ \ka} \equiv {1\over{\bbx}} \cc \lrdd \cc
 \hh {\bf P}\ll\chi + \hh {\bf P}\ll{\bar{\chi}} - {1\over 4} \cdot {\bf 1} \rrdd \ .
\end{align}
It is not difficult to deduce the form of the operator $\ka$. It is an R-neutral and Lorentz-scalar
operator, and taking the trace is an $R$-neutral and Lorentz-scalar operation, as is
taking the integrand of the supertrace and evaluating at zero.  On any eigenspace of
$P\ll\m$, there are exactly sixteen linearly independent operators anticommuting with
the $Q\ll \a$ and the $\bar{Q}\ll \ald$ which consist of 
all the operators composed of $D\ll\a$ and $\DDb\ll\ald$ that treat $P\ll\m$ as a $c$-number.  
Of those sixteen operators, only five are Lorentz-scalar, and of those 
only three are neutral under the R-symmetry: $1, \mathbf{P}\ll \chi$ and $\mathbf{P}\ll{\bar{\chi}}$.  
So combinations of these are the only operators that we can write as
arguments for ${\tt Istr}$ at $\th = \bar{\th} = 0$
for fixed momentum, and they are also the only operators that can have nonvanishing
traces with $\ka$.  There remain only three undetermined coefficients to fix,
and we fixed them.

Since $\ka$ commutes with $\bigp$ and all the operators that contribute to 
${\cal O}$ -- {\it i.e.} \mbox{$P\sqd = - \bbx$},  ${\bf P}\ll\chi$, ${\bf P}\ll{\bar{\chi}}$, 
and the identity -- we obtain
\begin{align}
{\tt Istr}_{E_P\otimes{\cal V}|0} \big(
\bigp \cc {\rm ln}({\cal O})
\big)
&= {\rm tr}_{E_P\otimes{\cal V}|0}\big(
\bigp \cc \ka\cc {\rm ln}({\cal O})\big)
\nonumber\\
&= {\rm tr}_{E_P\otimes{\cal V}|0}  \lrdd\bigp\,
{1\over\bbx} \cc \lrdd \cc \hh {\bf P}\ll\chi + \hh \cc {\bf P}
\ll{\bar{\chi}} - {1\over 4} \cdot {\bf 1}  \rrdd  {\rm ln} \big(
{\cal O} \big)
 \rrdd
\nonumber\\
&= {\rm tr}_{E_P\otimes{\cal V}|0} \cc \lrdd 
{1\over 4 } \cc {1\over\bbx} 
\bigp \cc {\rm ln}({\cal O}) \cc \rrdd
\\
&=-{1\over{4 P^2}}  {\rm ln} \cc {\tt Cdet}_{E_P\otimes{\cal V}|0} ({\cal O})\,.\nonumber
\end{align}
We define the spectral polynomial analogously to the single-field case 
\rr{SpectralPolynomial} as
\be\label{SpectralPolynomialDefinitionRecap}
\s(P\sqd) \equiv {\tt Cdet}_{E_P\otimes{\cal V}}\lrdd 
\sqrt{\bbx}\cc {\cal O}
\rrdd = \big(
-P^2 \big)
\uu {4 n} \cc {\tt Cdet}_{E_P\otimes{\cal V}}
({\cal O}) \ .
\ee
The relation between ${\tt Istr}\big(
\bigp \cc {\rm ln}({\cal O})
\big)$
evaluated at \mbox{$\th = \thb = 0$} and the spectral polynomial is then
\be\label{relation}
{\tt Istr}_{E_P\otimes{\cal V}|0}\big(
\bigp \cc {\rm ln}({\cal O})\big)
=  -{1\over{4P^2}} \cc {\rm ln}(\s(P\sqd)) + {n\over {P^2}} \cc {\rm ln}(-P^2)\ .
\ee
Since the spectral polynomial is again of
the form \eqref{sigmadef} with
\be
N\uu{1\over 4}  \cdot \prod\ll{j = 1}\uu J \m\ll j \sqd = \s(0)\uu{1\over 4}
 = |{\tt det}\ll{n\times n} (A_1)|^2\bigg|_{P\sqd = 0}\,,
\ee
following from equation~\eqref{multifieldbigplogcaloval}, this shows that in the case of several chiral superfields the one-loop effective K\"ahler potential also takes the form~\eqref{lastformula}.\\


\section{Examples}\label{secfive}

In this section we apply our formula in certain examples with the aim of illuminating
the regularization-dependence of the formula, and in particular the sense in which
the formula decouples the unphysical super-cutoff modes consistently despite 
treating them on a democratic footing with the physical modes, at the level of the momentum
integrand.

\setcounter{equation}{0}

\subsection{Higher derivative example}

\heading{Computing the one-loop K\"ahler potential with the mass-dependent formula}
Consider a simple Wess-Zumino model with a $\Box/M^2$ type operator.  Let the Lagrangian to quadratic order in fluctuations be
\be
\L=\int d^2\theta\, d^2\bar\theta\,\chi^\dagger\left(1-\frac\Box{|M|^2}\right)\chi+\int d^2\theta\,\frac12\chi W''\chi+\int d^2\bar\theta\, \frac12\chi^\dagger\overline W''\chi^\dagger
\ee
From \eqref{FinalFormula} with $A_2=1-\Box/|M|^2$ and $A_1=W''$, we can immediately write down the one-loop correction to the K\"ahler potential.
\be
\Delta K=\frac\hbar{2(2\pi)^4}\int d^4P\frac1{P^2}\ln\left[\left(1+\frac{P^2}{|M|^2}\right)^2+\frac{|W''|^2}{P^2}\right]\,.
\ee
It is perhaps not obvious that the zeroes of the argument of the log still give masses, but this becomes apparent upon considering the equations of motion,
\be\label{eom}
\left(1-\frac\Box{|M|^2}\right)\chi=\frac{\overline W''\bar D^2}{4\Box}\chi^\dagger\,.
\ee
By acting with $D^2$ on \eqref{eom} and combining the result with the conjugate equation, we find
\be
\left(1-\frac{\Box}{|M|^2}\right)^2=\frac{|W''|^2}\Box\,,
\ee
the solutions of which are
\begin{align}
\mu_L^2&=|W''|^2\left(1+2\left|\frac{W''}{M}\right|^2+7\left|\frac{W''}{M}\right|^4+{\cal O}\left(\left|\frac{W''}{M}\right|^6\right)\right)\,,\\
\mu_{H\pm}^2&=|M|^2\left(1\pm\left|\frac {W''}{M}\right|-\frac{1}{2}\left|\frac {W''}{M}\right|^2+{\cal O}\left(\left|\frac{W''}{M}\right|^3\right)\right)\,.
\end{align}
\heading{Direct computation by expanding the action in ${1\over M\sqd}$}
For comparison, consider now the calculation of the (1-PI) effective action for
a model with action
\be
S[\Phi,\Phi^\dagger]=\int d^8z\,\Phi^\dagger\left(1-\frac\Box{|M|^2}\right)\Phi+\int d^6z\,W(\Phi)+\int d^6\bar{z}\,\overline W(\Phi^\dagger)
\ee
in dimensional regularization to subleading order in $1/M$. The action as written is assumed to define an effective field theory that describes the light degrees of freedom below some matching scale $\mu< M$. One can then simply treat the higher derivative term perturbatively. 

The goal will then be to extract the correction to the K\"ahler potential from
\begin{align}
\Gamma[\Phi,\Phi^\dagger]&=S[\Phi,\Phi^\dagger]\\\nonumber
&\,-i\ln\!\left[
\int\!\mathcal{D}\chi\mathcal{D}\chi^\dagger 
\text{exp}\left\{
i\!\!\int \!d^8z \chi^\dagger\!\left(1-\frac\Box{|M|^2}\right)\!\chi\!+i\!\!\int\! d^6z\frac12\chi W''\chi\!+i\!\!\int\! d^6\bar z \frac12\chi^\dagger\overline W''\!\chi^\dagger
\right\}\right]\!,
\end{align}
treating the higher derivative term as a perturbation. This gives
\begin{align}
\Gamma[\Phi,\Phi^\dagger]&=S[\Phi,\Phi^\dagger]+\frac{i}{2}\str \bigp\ln\mathcal{O}-i\left[-i\int d^8z \langle\chi^\dagger(z)\left(\frac\Box{|M|^2}\right)\chi(z)\rangle_c\right.\nonumber\\
&\left.+\frac{(-i)^2}{2}\int d^8z\int d^8z'\langle\chi^\dagger(z)\left(\frac\Box{|M|^2}\right)\chi(z)\chi^\dagger(z')\left(\frac\Box{|M|^2}\right)\chi(z')\rangle_c+\cdots\right]
\end{align}
where $\mathcal{O}$ has $A_1=W''(\Phi)$ and $A_2=1$ and the subscript indicates that only the connected contributions are to be kept. To extract the correction to the K\"ahler potential, as before we evaluate at $\Phi=\phi$, $\Phi^\dagger=\phi^*$. To subleading order in the interaction, we then have
\begin{align}
\Delta K(\phi,\phi^*)&=\frac{i}{2V}{\tt Istr} \bigp\ln\mathcal{O}|_{\theta=\bar\theta=0}\nonumber\\
&-\frac{1}{|M|^2}\int d^8z'\,\delta^8(z-z')|_{\theta=\bar\theta=0}\Box\langle\chi(z)\chi^\dagger(z')\rangle|_{\theta=\bar\theta=0}\nonumber\\
&+\frac{i}{2|M|^4}\int d^8z'\Box\langle\chi(z)\chi^\dagger(z')\rangle|_{\theta=\bar\theta=0}\Box\langle\chi(z')\chi^\dagger(z)\rangle|_{\theta=\bar\theta=0}\nonumber\\
&+\frac{i}{2|M|^4}\int d^8z'\Box\Box'\langle\chi(z)\chi(z')\rangle|_{\theta=\bar\theta=0}\langle\chi^\dagger(z')\chi^\dagger(z)\rangle|_{\theta=\bar\theta=0}\nonumber\\
&+\dots\,.
\end{align}
The first term is the usual correction for the renormalizable theory. So we focus on corrections suppressed by $M$. Using the explicit form for the propagator in this background
\begin{align}
\langle\chi(z)\chi^\dagger(z')\rangle&=i\frac{\overline{D}^2D^2}{16(\Box-|m|^2)}\delta^8(z-z')\,,\nonumber\\
\langle\chi(z)\chi(z')\rangle&=i\frac{m^*\overline{D}^2}{4(\Box-|m|^2)}\delta^8(z-z')\,,\nonumber\\
\langle\chi^\dagger(z)\chi^\dagger(z')\rangle&=i\frac{m D^2}{4(\Box-|m|^2)}\delta^8(z-z')\,,
\end{align}
where $m=W''(\phi)$, one finds (after continuation to Euclidean momenta)
\begin{align}
\Delta K(\phi,\phi^*)\supset\,\,&\frac{1}{|M|^2}\int \frac{d^4P}{(2\pi)^4}\left(1-\frac{|m|^2}{P^2+|m|^2}\right)\nonumber\\
&-\frac{1}{2|M|^4}\int \frac{d^4P}{(2\pi)^4}\frac{P^6}{(P^2+|m|^2)^2}\nonumber\\\,
&+\frac{1}{2|M|^4}\int \frac{d^4P}{(2\pi)^4}\frac{|m|^2P^4}{(P^2+|m|^2)^2}\nonumber\\\,
&+\cdots\,.
\end{align}
Performing the integral in $4-2\epsilon$ dimensions and working with the $\overline{\text{DR}}$-scheme, we find
\be
\Delta K(\phi,\phi^*)\supset\frac{|m|^4(1-\ln |m|^2/\mu^2)}{16\pi^2|M|^2}+\frac{|m|^6(5-7\ln |m|^2/\mu^2)}{32\pi^2|M|^4}+\cdots\,.
\ee
Recalling that $m=W''(\phi)$, we finally have
\be
\Delta K(\Phi,\Phi^\dagger)\supset\frac{|W''(\Phi)|^4(1-\ln |W''(\Phi)|^2/\mu^2)}{16\pi^2|M|^2}\\ +\frac{|W''(\Phi)|^6(5-7\ln |W''(\Phi)|^2/\mu^2)}{32\pi^2|M|^4}+\cdots\,.
\ee
This agrees with our formula provided only the light mode is kept. Performing the integrals with a Wilsonian cut-off instead, one finds agreement with the formula if all modes are kept. The reason the heavy modes do not contribute to the final answer in perturbation theory in dimensional regularization can be traced to the fact that the large $M$ expansion of their logarithms leads to scale-free integrals.

That is, in dimensional regularization the formula treats the unphysical super-cutoff and physical modes on a democratic footing at the level of the integrand prior to expanding in the
heavy scale $M$.  It is only this expansion that breaks the symmetry among modes that
is present until this point.  For a Wilsonian regulator, physical and super-cutoff modes
both make nonzero contributions to the momentum integrand, but the momentum-dependent contributions of the supercutoff modes are small, with a power of $M\sqd$ appearing automatically in the denominator for each power of $P\sqd$ in the numerator.
Supercutoff modes thus contribute to the integral with powers $\Lambda\sqd / M\sqd$, since our domain of integration
is restricted to $|P| < \Lambda$.

For a Wilsonian cutoff, the processes of expanding the action, expanding the integrand, and doing
the integral, all commute with each other.  For dimensional regularization,
the first two commute and the third does not.
Since the first two processes commute for either regulator, terms of order $M\uu{-(m+1)}$ and 
smaller in the treel-level action, manifestly have no effect on the one-loop value of
$\Delta K$ up to order $M\uu{-m}$.

\subsection{The two-dimensional sigma model}
Let us now check our formula against the known expression \cite{Friedan:1980jm} 
for the $\beta$-function
in the case of two-dimensional 
sigma models.  Although our formula depends only on physical
 masses, we shall find that by taking a limit as the masses go to zero, we 
 recover the correct $\beta$-function governing the anomalous scale dependence
 of the massless limit.
 
In going from four to two space-time dimensions, $\mathcal{N}=1$ supersymmetry 
reduces to (2,2) supersymmetry. A K\"ahler potential which is a general
polynomial in chiral fields yields a two-dimensional $\mathcal{N}=(2,2)$ sigma
model where the scalar components of the fields parametrize a K\"ahler 
manifold.  For the action
\begin{equation}\label{sigmamodelaction}
S=\int d^2x\,d^4\theta\,K(\Phi^i,\Phi^{\dagger\bar{j}})\,,
\end{equation}
we verify that \eqref{lastformula}
reproduces the well-known fact that the one-loop beta function of
the K\"ahler metric $K_{i\bar{j}}=\partial/\partial\phi^i\,\partial/\partial\phi^{\bar{j}*}\,K$  
as a coupling is proportional to the Ricci curvature \cite{Friedan:1980jm}. We shall do this in
a Wilsonian regularization scheme, where for a momentum-space UV cut-off
$\Lambda$
\begin{equation}
\beta^{K_{i\bar{j}}} \equiv -\Lambda\frac{\delta}{\delta\Lambda}K_{i\bar{j}}(\Lambda)\ .
\end{equation}
As \eqref{lastformula}
holds in a theory with massive linearized quantum fluctuations,
we must first add masses to our sigma model. This is done by adding a 
superpotential term to the action \eqref{sigmamodelaction},
\begin{equation}\label{sigmamodelaction2}
S=\int d^2x\,d^4\theta\,K(\Phi^i,\Phi^{\dagger\bar{j}})+\left[\int d^2x\,d^2\theta\,W(\Phi^i)\,+{\rm h.c.} \right]\,,
\end{equation}
and computing the mass-squared matrix of the linearized modes
in the background field formalism, before eventually 
taking the zero-mass limit. Hence we apply the splitting
\eqref{simpleansatzmemo}
and consider the quadratic Lagrangian.
Integrating out the auxiliary components of the fluctuations
and performing the integration over the Grassmann parameters leaves us with
\begin{equation}\label{LagrangianinSigmaModel}
\mathcal{L}=K_{i\bar{j}}\left[
\partial_{\mu}\chi_0^i\partial^{\mu}\chi_0^{\bar{j}\,*}-K^{\bar{k}i}\overline{W}_{\bar{k}\bar{l}}K^{\bar{j}m}W_{mn}\chi_0^{\bar{l}\,*}\chi_0^n
\right]+\ldots\,,
\end{equation}
where $\chi_0$ $(\chi_0^*)$ are the scalar  components of the fluctuation superfields $\chi$ ($\bar{\chi}$),
$K_{i\bar{j}}\equiv K_{i\bar{j}}(\phi,\phi^*)$ is the K\"ahler potential evaluated on the constant background fields
(with inverse matrix $K^{\bar{j}i}$), and $W_{ij}=\partial/\partial\phi^i\,\partial/\partial\phi^jW(\phi)$ 
are the second derivatives of the superpotential evaluated in the background.
The omitted terms contain only the fermionic components of the fluctuations, such that
we can read off the mass-squared matrix from the equations of motion for the $\chi_0^i$ as obtained from
\eqref{LagrangianinSigmaModel}
\begin{equation}
\partial^{\mu}\partial_{\mu}\chi_0^i=-M^i_{\;j}\,\chi_0^j\,,
\end{equation}
with
\begin{equation}\label{sigmamodelmassmatrix}
M^i_{\;j}=K^{\bar{k}i}\,\overline{W}_{\bar{k}\bar{l}}\,K^{\bar{l}m}\,W_{mj}\equiv({\bf K}^{-1})^T\,\overline{\bf W}\,{\bf K}^{-1}\,{\bf W}\,,
\end{equation}
in an obvious notation.
The hermitian conjugate fields lead an analogous matrix which we combine with
\eqref{sigmamodelmassmatrix} into one mass-squared matrix $M^I_{\,J}$, where
$I$ and $J$ run over both $i$ and $\bar{j}$ indices.
For the process of changing our cut-off from $\Lambda$ to $\Lambda-\Delta\Lambda$ by integrating out modes, 
\eqref{lastformula} 
yields
\begin{align}
\Delta K =&\frac{1}{4\pi}\int_\Lambda^{\Lambda-\Delta\Lambda}\frac{dP}{P}\ln\left[\prod_{I}\frac{P^2+\mu_I^2}{\mu_I^2}\right]\\
=&\,\frac{\Delta\Lambda}{4\pi\Lambda}\,\left[\ln\left(\prod_I\mu^2_I\right)-\ln\left(\prod_I(\Lambda^2+\mu_I^2)\right)\right]+\ldots\,,\nonumber
\end{align}
where we have set $\hbar=1$ and omitted higher-order terms in $\Delta\Lambda$ in the second line.
The $\mu_I^2$ are the eigenvalues of the mass-squared matrix ${\bf M}$, which we can
split according to \eqref{sigmamodelmassmatrix} such that we obtain
\begin{align}
\Delta K &=\,\frac{\Delta \Lambda}{4\pi\Lambda}\left[\ln\det {\bf M}-\ln\det({\Lambda\sqd\cdot\bf 1}+{\bf M})\right]\nonumber\\
&=\,-\frac{\Delta \Lambda}{4\pi\Lambda}\bigg[
\ln\det |{\bf K}|^2+\ln\det({\bf 1}+\Lambda^{-2}{\bf M})\\
&\qquad\qquad\quad-\ln\det{\bf W}-\ln\det\overline{{\bf W}}
+\ln(\Lambda\sqd)
\bigg]\,,\nonumber
\end{align}
again up to higher orders in $\Delta\Lambda$.
The last three terms in the square bracket of the second equation are either holomorphic 
or antiholomorphic and can be discarded. If we then take the superpotential to 
zero, the mass-squared matrix vanishes and the remaining terms have a smooth limit.  By taking two derivatives with respect to the 
zero modes of the scalar fields and using an identity in K\"ahler geometry,
\begin{equation}
R_{i\bar{j}} = - \partial_i \partial_{\bar{j}} \ln\det \bf K\,,
\end{equation}
 we obtain the familiar result for the one-loop $\beta$-function
for the supersymmetric nonlinar sigma model:
\begin{equation}
\beta^{K_{i\bar{j}}}=\frac{1}{2\pi}\frac{\partial}{\partial\phi^i}\frac{\partial}{\partial\phi^{\bar{j}*}}\ln\det {\bf K} =-\frac{1}{2\pi}R_{i\bar{j}}\,.
\end{equation}

\section{Conclusions and Outlook}

In this paper we have proven a universal formula for the one-loop renormalization of
the K\"ahler potential due to chiral multiplets circulating in the loop.  The formula 
generalizes previous results
 \cite{Grisaru:1996ve},\cite{Brignole:2000kg} in that it applies to general effective actions
 with arbitrary numbers of derivatives.  It would be useful to see to what extent a similar mass-dependent formula applies to the contribution to the one-loop K\"ahler potential of multiplets other than chiral multiplets.  In particular, the contributions of massive vector multiplets are relevant for the study of realistic theories.  For vector multiplets, there are delicate questions of gauge dependence of the K\"ahler potential off the D-flat moduli space, for which reason one must be careful about defining the effective K\"ahler potential in a
gauge-invariant way.  We remark here that
a straightforward calculation can be done in unitary gauge \cite{ToAppear}, 
precisely in parallel with the calculation presented in this paper,
with the result that the massive vector multiplet contributes precisely in the same form as a chiral multiplet of the same mass, with an overall factor of -2.  This value is in manifest agreement with \cite{Grisaru:1996ve} in the renormalizable case, and somewhat
nontrivially in agreement with \cite{Brignole:2000kg} in the non-renormalizable two-derivative case.
 
Four-dimensional renormalizable theories are formulated in terms of chiral and vector
multiplets alone, but since our formula applies to effective field theories with a finite cutoff, there is
nothing in principle to obstruct extending the analysis presented here to theories with massive higher-spin multiplets.  To extend the formula to perturbative but non-Lagrangian theories
such as superstring theory would also be interesting, and potentially of value in
model building.

\section*{Acknowledgments}
\addcontentsline{toc}{section}{Acknowledgments}
We would like to thank several people for valuable discussions, in particular Zohar Komargodski, Massimo Porrati, Taizan Watari, and also the IPMU joint string/phenomenology group meeting for some useful early comments and suggestions.  The work of S.H., C.S.-C., and M.S. was supported by the World Premier International Research Center Initiative, MEXT, Japan.  The work of S.H. was also supported by a Grant-in-Aid for Scientific Research (22740153) from the Japan Society for Promotion of Science (JSPS). The work of R.F. has been partially supported by the National Science Foundation under Grant No. NSF-PHY-0855425.  R.F. would like to thank IPMU for hospitality while some of the results presented here were obtained.


\section*{Appendices}
\addcontentsline{toc}{section}{Appendices}
\setcounter{section}{0}
\renewcommand{\thesection}{\Alph{section}}
\renewcommand{\theequation}{\Alph{section}\arabic{equation}}

\section{Superspace conventions}\label{AppendixWBconventions}

\bi
\item In all our text, $\bbx \equiv - \ppp\ll t \sqd + \ppp\ll i \sqd = \eta\uu{\m\n}
\ppp\ll\m \ppp\ll\n$, where $\eta\uu{\m\n}$ is the mostly-plus metric.
\item For $N=1$ superspace coordinates $(x,\theta,\bar{\theta})$ in $D=4$
dimensions we have
\begin{equation}
\begin{array}{ll}
Q=\frac{\partial}{\partial\theta}-i\sigma^m\bar{\theta}\partial_m\,, & \bar{Q}=-\frac{\partial}{\partial\bar{\theta}}+i\theta\sigma^m\partial_m\,,\\
D=\frac{\partial}{\partial\theta}+i\sigma^m\bar{\theta}\partial_m\,, \qquad& \bar{D}=-\frac{\partial}{\partial\bar{\theta}}-i\theta\sigma^m\partial_m\,,
\end{array}
\end{equation}
where $\partial_m$ is a spatial derivative, $\sigma^m$ is the vector consisting of the Pauli matrices and the identity,
and spinor indices $\alpha,\,\dot{\alpha}$
are suppressed. If we apply these operators on a chiral superfield, we sometimes
implicitly assume that the space-time coordinates have been shifted
$x\mapsto x+i\theta\sigma\bar{\theta}$, in which case the expressions
are rather
\begin{equation}
\begin{array}{ll}
Q=\frac{\partial}{\partial\theta}\,, & \bar{Q}=-\frac{\partial}{\partial\bar{\theta}}+2i\theta\sigma^m\partial_m\,,\\
D=\frac{\partial}{\partial\theta}+2i\sigma^m\bar{\theta}\partial_m\,, \qquad& \bar{D}=-\frac{\partial}{\partial\bar{\theta}}\,.
\end{array}
\end{equation}
The derivatives satisfy in particular the identities
\be\label{Drivativerelations}
 D\ssqd \DDb\ssqd D\ssqd = + 16 \bbx D\ssqd \ , \llsk
\DDb\ssqd D\ssqd \DDb\ssqd = + 16 \bbx  \DDb\ssqd \ ,
\ee
and
\be\label{IstrD}
\istrz( \DDb \ssqd D\ssqd ) = \istrz( D \ssqd \DDb\ssqd ) = 16\,.
\ee

\item The operator
\be\label{pchdefmemo}
\pch\equiv {{\DDb\ssqd D\ssqd}\over{16 \bbx}}\ , \llsk \pch\sqd = \pch\ ,
\ee
is a projection onto functions on superspace satisfying the chiral constraint.
Likewise
\be\label{pachdefmemo}
\pach\equiv {{D\ssqd \DDb\ssqd }\over{16 \bbx}}\ , \llsk \pach\sqd = \pach\ ,
\ee is a projection operator onto
functions on superspace satisfying the antichiral constraint.
These two projectors are mutually excluding,
\be
\pch \pach = \pach \pch = 0\ ,
\ee
but not complementary,
\be\label{prsdefmemo}
\prs\equiv \pch + \pach \neq 1\ .
\ee
We define the complement of $\prs$ to be
\be\label{prtdefmemo}
\prt \equiv 1 - \prs\ ,
\ee
which projects onto the
``transverse part" or ``gauge invariant part" of a real superfield.
For the projectors $\pch, \pach$ and the derivatives $D$, $\bar{D}$ we have
the following identities:
\begin{align}\label{projidentsmemo}
\pch \cc D\ssqd &= \DDb\ssqd \cc \pch = \pach\cc \DDb\ssqd = D\ssqd \cc \pach
= 0\,,\nonumber\\
\pach\cc D\ssqd &= D\ssqd \cc\pch = D\ssqd\ , \llsk \pch\cc \DDb\ssqd
= \DDb\ssqd \cc \pach = \DDb\ssqd\ .
\end{align}
Furthermore, we have
\be\label{projectortraces}
{\rm tr}\ll{\cal V}(\pch) = {\rm tr}\ll{\cal V} (\pach) = 4\ ,\qquad
{\rm tr}\ll{\cal V}(\prs) = {\rm tr}\ll{\cal V} (\prt) = 8\ ,
\ee
and
\be\label{UsefulFormulaA}
\istr_{{\cal V}|0} (\pch) = \istr_{{\cal V}|0}(\pach) =  {1\over \bbx}\ ,\qquad
\istr_{{\cal V}|0}(\bigp) =  {2\over\bbx}\ .
\ee

\ei

\section{The kinetic operator ${\cal O}$, its logarithm, and the spectral polynomial}\label{AppendixOproperties}

In this appendix we calculate the logarithm of ${\cal O}$, and its product with the projector
$\bigp$.  We begin by representing ${\cal O}$ in terms of a matrix $B$
that squares to a matrix proportional to $\bigp$.  Defining
\be\label{bigBdef}
B\equiv \left (\bm  0 & - {{\DDb\ssqd}\over {4\bbx}  }  \cc {{A\ll 1 \st (\bbx)}\over{
A\ll 2 (\bbx)}} \cr
 - {{D\ssqd}\over {4\bbx}  }  \cc {{A\ll 1 (\bbx)}\over{A\ll 2 (\bbx)}}  & 0 \em \right ) \  ,
\ee
we have
\be
B \cc \bigp = \bigp \cc B\ ,
\ee
\be
{\rm tr}\ll{2\times 2} \cc (B) =  {\rm tr}\ll{2 \times 2}\cc (\bigp \cc B) = 0\ ,
\ee
\be
{\cal O}  =  A\ll 2 \cc (1 + B)\ ,
\ee
and
\be
B\sqd = {{|A\ll 1|\sqd}\over{\bbx\cc A\ll 2\sqd}}\cc \bigp\ ,    
\ee
where we have used the definitions \rr{pchdefmemo} and \rr{pachdefmemo}
and the properties \rr{projidentsmemo}. 
We will always be interested in tracing ${\rm ln}({\cal O})$ or
its product with $\bigp$ as a $2\times 2$ matrix, so odd powers of $B$ will
never be of interest to us, as they have zeroes on the diagonal.  Thus
we can write
\begin{align}
{\rm ln}(1 + B) &= ({\rm traceless~}2\times 2) + \hh {\rm ln}(1 - B\sqd)
\nonumber\\
&= ({\rm traceless~}2\times 2) + \hh  
{\rm ln}(1 - {{|A\ll 1|\sqd}\over{\bbx\cc A\ll 2\sqd}} \cc \bigp ) 
\nonumber\\
&= ({\rm traceless~}2\times 2) + \hh  \bigp\cc
{\rm ln}(1 - {{|A\ll 1|\sqd}\over{\bbx\cc A\ll 2\sqd}}  ) \ ,
\end{align}
and
\be
\bigp \cc {\rm ln}(1 + B)
= ({\rm traceless~}2\times 2) + \hh  \bigp\cc
{\rm ln}(1 - {{|A\ll 1|\sqd}\over{\bbx\cc A\ll 2\sqd}}) \ ,
\ee
so
\be
{\rm ln}({\cal O} ) = {\rm ln}(A\ll 2) \cc {\bf 1} + 
 \hh  
\bigp\cc {\rm ln}(1 - {{|A\ll 1|\sqd}\over{\bbx\cc A\ll 2\sqd}} ) 
+  ({\rm traceless~}2\times 2) \ ,
\ee
and
\begin{align}\label{bigplogcaloval}
\bigp \cc {\rm ln}({{\cal O}}) &= \bigp \cc \lsqq \cc {\rm ln}(A\ll 2) + \hh 
{\rm ln}(1 - {{|A\ll 1|\sqd}\over{\bbx\cc A\ll 2\sqd}} )  \cc\rsqq
+  ({\rm traceless~}2\times 2) 
\nonumber\\
&= \hh \cc \bigp \cc {\rm ln} \lrdd  A\ll 2\sqd - {{|A\ll 1|\sqd}\over{\bbx}} \cc \rrdd \cc
+  ({\rm traceless~}2\times 2) 
\end{align}
Together with \rr{UsefulFormulaA}, we use this identity to derive formula \rr{FinalFormula}.\\

We also use the matrix $B$ to compute the logarithm of the spectral 
polynomial $\s(P\sqd)$. This is given by
\begin{align}
{\rm ln}(\s(P\sqd)) =
{\tt Ctr}\left(
{\rm ln}\big(
\sqrt{\bbx}\cc \co\big)
\right)
 &=
{\tt Ctr}\left(
{\rm ln}\big(
\sqrt{\bbx}\cc A\ll 2(\bbx)\big)
\cdot{\bf 1} \right)
+ {\tt Ctr}\left(
{\rm ln}\big(
{\bf 1} + B\big)
\right)
\nonumber\\
&=  {\tt tr}(\bigp) \cdot \left(
{\rm ln}\big(
\sqrt{\bbx}\cdot A\ll 2(\bbx)\big)
+ \hh\cc {\rm ln}\big(
1 - {{|A\ll 1(\bbx)|\sqd }\over{\bbx\cc A\ll 2(\bbx)\sqd}}\big)
\right)
\nonumber\\
&= \hh\cc {\tt tr}(\bigp) \cdot {\rm ln}\left( 
\bbx\cdot\big(
A\ll 2(\bbx)\sqd - 
\tfrac{|A\ll 1(\bbx)|\sqd}{\bbx} \big)
\right)\ .
\end{align}
Now, ${\tt tr}(\bigp) = {\tt Ctr}({\bf 1}) = 8$, so
\be
{\rm ln}(\s(P\sqd)) = 4\cc {\rm ln} \left(
\bbx\cdot\big(
A\ll 2(\bbx)\sqd - \tfrac{|A\ll 1(\bbx)|\sqd}{\bbx} \big)
\right)\ ,
\ee
and thus
\be\label{logsigmaformula}
\s(P\sqd) = \bbx^4 \cc \big(
A\ll 2(\bbx)^4 - \tfrac{|A\ll 1(\bbx)|\sqd}{\bbx}\big)
^4
\ee
For formula \eqref{relation}, we also need that
\begin{align}\label{additional}
{1\over{4\cc\bbx}}  {\rm ln} \cc {\tt Cdet}\ll{E_P} ({\cal O})
&= {1\over{4\bbx}} \cc 
{\rm ln} \lrdd{ {{\tt Cdet} \cc (\sqrt{\bbx} \cc{\cal O})}
\over{{\tt Cdet}(\sqrt{\bbx}\cdot{\bf 1})}}  \rrdd 
\nonumber\\
&= {1\over{4\cc\bbx}} \cc \left(
{\rm ln}(\cc \s(P\sqd)) - 
{\rm ln} \cc {\tt Cdet} \cc (\sqrt{\bbx} \cdot{\bf 1})\right)
\\
&= {1\over{4\bbx}} \cc {\rm ln}(\s(P\sqd)) - {1\over{4\bbx}} \cc {\rm ln} \left(
\big(\sqrt{\bbx}\big)\uu {8}\right)
\,.\nonumber
\end{align}

In section \ref{sec:multifield}, we want to perform a similar computation
in the case where we have several chiral superfields. 
It is useful to use the  following identities for determinants of block matrices:
\begin{align}
{\rm det} \cc \lrdd \cc \begin{matrix} A & c \cr b & D \end{matrix} \cc \rrdd
&= {\rm det}(D) \cc {\rm det} \left(
A - c \cc D\uu{-1} \cc b \right)
\nonumber\\
&= {\rm det}(A) \cc {\rm det}(D) \cc {\rm det} \left(
1 - A\uu{-1} \cc c \cc D\uu{-1} \cc b \right)\ .
\end{align}
In our case we have $A=D\st=A_2(\bbx),\; b = c\st = - {{D\sqd}\over{4\bbx}} \cc A_1(\bbx)$.  Then for
constant, supersymmetric backgrounds the operators $D, \bar{D}$ and $\bbx$
all commute through $A_1, \, A_2$ and their conjugates, so
\be
{\rm det}({\cal O}) = 
\lba\cc {\rm det}(A_2) \cc \rba \sqd \cc {\rm det} \lrdd 
1 - A_2\uu{-1}\cc  A_1 \st \cc A_2\uu{*-1}
\cc A_1\cc {{P\ll\chi}\over{\bbx}} \rrdd\ .
\ee

This is not yet quite  the object we want; we really want a constrained determinant
rather than a full determinant.  
We have
\be
{\tt Cdet}\lrdd \cc \begin{matrix} A & c \cr b & D \end{matrix} \cc \rrdd = 
{\tt Cdet} \cc\lrdd \begin{matrix} A & 0 \cr 0 & D \end{matrix} \rrdd \;
{\tt Cdet} \cc (1 + K) \ ,
\ee
where
\be
K \equiv \lrdd \cc \begin{matrix}  0 &  A\uu{-1} \cc c \cr D\uu{-1} \cc b & 0 \end{matrix} \cc \rrdd\ . 
\ee
For any $a \geq 1$,
the matrix $K$ has the properties that $K\uu a = \bigp \cc K\uu a = K\uu a \cc \bigp$
and ${\tt Ctr}(K\uu {2 a - 1}) = 0$.  For any matrix with these properties,
\be
{\tt Cdet}(1 + K) =  \sqrt{\cc {\tt det} \cc  (1 - K\sqd) }\ . 
\ee
Here, 
\be
K\sqd = \lrdd \cc \begin{matrix} A\uu{-1} c D\uu{-1} b & 0 \cr 0 & D\uu{-1} b A\uu{-1} c 
\end{matrix} \cc \rrdd\ .
\ee
For us, \eqref{Omultifield} leads to
\be
K\sqd =  \lrdd \cc \begin{matrix} A_2\uu{-1} A_1\uu* A_2\uu{*-1} A_1 \cc {{P\ll\chi}\over \bbx}
 & 0 \cr 0 &  A_2\uu{*-1} A_1\st A_2\uu{-1} A_1\st \cc {{P\ll{\bar{\chi}}}\over \bbx}  \end{matrix} \cc  \rrdd\,.
\ee
Then
\be
{\tt det} \cc  (1 - K\sqd) =
\left(
 {\tt det}\ll{n\times n} (1 - A_2\uu{-1} A_1\st A_2\uu{*-1} A_1\cc\bbx\uu{-1})\right)\uu 8
\ee
and
\begin{align}
{\tt Cdet}(1 + K) &=\left(
 {\tt det}\ll{n\times n} (1 - A_2\uu{-1}A_1 \st A_2\uu{*-1} A_1 \cc\bbx\uu{-1} ) \right)\uu 4
 \nonumber\\&=
\left(
 {\tt det}\ll{n\times n} (1 - A_2\uu{*-1} A_1  A_2\uu{-1} A_1\st \cc\bbx\uu{-1} ) \right)\uu 4 
\end{align}
and
\be
{\tt Cdet} \cc\lrdd \begin{matrix} A & 0 \cr 0 & D \end{matrix} \rrdd
=
{\tt Cdet} \cc\lrdd \begin{matrix} A_2 & 0 \cr 0 & A_2\st \end{matrix} \rrdd
=
 \left( {\tt det}\ll{n\times n}(A_2)\right)\uu 4\cc
\left( {\tt det}\ll{n\times n}(A_2\st)\right)\uu 4\ ,
\ee
so
\begin{align}\label{multifieldbigplogcaloval}
{\tt Cdet}({\cal O}) &= \lba \cc {\tt det}\ll{n\times n} (A_2) \cc \rba\uu 8 \cc
\left(
 {\tt det}\ll{n\times n} (1 - A_2\uu{*-1} A_1  A_2\uu{-1} A_1\st \cc\bbx\uu{-1} ) \right)\uu 4 
\nonumber\\
&=  \left(
 {\tt det}\ll{n\times n} (A_2\cc A_2\st - A_1  A_2\uu{-1} A_1\st A_2 \cc \bbx\uu{-1} )\right)\uu 4 \ .
\end{align}


\end{CJK*}\end{document}